# Unexpected versatile electrical transport behaviors of ferromagnetic nickel films

Kai-Xuan Zhang[1*], Hanshu Xu[2], Jihoon Keum[1], Xiangqi Wang[3], Meizhuang Liu[4], and Zuxin Chen[4,5*]

[1] Center for Quantum Materials, Department of Physics and Astronomy, Seoul National University, Seoul 08826, South Korea
[2] Department of Applied Physics, School of Biomedical Engineering, Anhui Medical University, Hefei 230032, China
[3] Jihua Laboratory Testing Center, Jihua Laboratory, Foshan 528000, China
[4] School of Semiconductor Science and Technology, South China Normal University, Foshan 528225, China
[5] Guangdong Provincial Key Laboratory of Chip and Integration Technology, Guangzhou 510631, China

E-mail: kxzhang@snu.ac.kr; chenzuxin@m.scnu.edu.cn

## Abstract

Perpendicular magnetic anisotropy (PMA) of magnets is paramount for electrically controlled spintronics due to their intrinsic potentials for higher memory density, scalability, thermal stability and endurance, surpassing an in-plane magnetic anisotropy (IMA). Nickel film is a long-lived fundamental element ferromagnet, yet its electrical transport behavior associated with magnetism has not been comprehensively studied, hindering corresponding spintronic applications exploiting nickel-based compounds. Here, we systematically investigate the highly versatile magnetism and corresponding transport behavior of nickel films. As the thickness reduces within the general thickness regime of a magnet layer for a memory device, the hardness of nickel films' ferromagnetic loop of anomalous Hall effect increases and then decreases, reflecting the magnetic transitions from IMA to PMA and back to IMA. Additionally, the square ferromagnetic loop changes from a hard to a soft one at rising temperatures, indicating a shift from PMA to IMA. Furthermore, we observe a butterfly magnetoresistance resulting from the anisotropic magnetoresistance effect, which evolves in conjunction with the thickness and temperature-dependent magnetic transformations as a complementary support. Our findings unveil the rich magnetic dynamics and most importantly settle down the most useful guiding information for current-driven spintronic applications based on nickel film: The hysteresis loop is squarest for the ~8 nm-thick nickel film, of highest hardness with $R_{xy}^r/R_{xy}^s$~1 and minimum $H_s$-$H_c$, up to 125 K; otherwise, extra care should be taken for a different thickness or at a higher temperature.

## 1. Introduction

Spintronics[1-4] has emerged as an engaging research field for a variety of magnetic materials, where the ability to electrically control a quantum magnetic state is essential for device applications. Basically, the key technology, *i.e.*, the current-driven magnetization switching can be achieved through spin-transfer torque[5-10] and the newly developed spin-orbit torque[11-21]. In both cases, the magnet layer is better to host a perpendicular magnetic anisotropy (PMA)[22] for practical devices[23], since this allows for higher areal density[24] and scalability[24-26], stronger thermal stability and endurance[24, 25], and lower power consumption[24, 25, 27] compared to in-plane magnetic anisotropy (IMA). Therefore, pursuing a suitable PMA is highly demanding for practical spintronic applications at a fundamental level.

Nickel film has long been considered a fundamental element magnet[28-30], hosting fascinating spin dynamics[29, 31] and promising future applications[29, 30, 32]. Gaining a comprehensive understanding of its magnetic behavior is crucial for the development of nickel-based ferromagnetic compounds toward spintronics. Surprisingly, research in this area is limited with no reports on systematic electrical transport investigation of nickel film, which is a basic if not least requirement for current-related spintronics. On the other hand, spin-orbit torque [11-17] is essential for next-generation spintronics, which exists in conventional ferromagnet/heavy-metal systems. Recently, material scientists of diverse backgrounds started to explore new materials for replacing the heavy-metal layer to achieve better performance, including high energy efficiency, magnetic field-free switching, etc. Before doing this, one should use any ferromagnet layer; in our case, the element ferromagnet nickel film for several reasons: cheap, nontoxic, relatively air-stable, and most importantly, easily reachable in any nanofabrication lab with evaporation or sputter machine. However, the systematic electrical transport study of nickel film is absent in the literature. Such missing guides on nickel film's transport are essential for the first-step optimization of ferromagnet's parameters, which motivates us to perform a comprehensive transport study on nickel film to lay down the research foundation for the next-step spin-orbit torque material screening. For this purpose, we have focused on studying the electrical magnetotransport response of nickel films below tens of nanometers thick, covering the common thickness range of ferromagnets in a practical magnetic memory device.

In this work, we investigate the electronic transport properties of nickel films at various thicknesses and temperatures, as well as the versatility of their magnetism. As the thickness of the film decreases successively from 24 to 8, 3, and 2 nm, we observe a gradual change from a residual resistance to a resistance upturn at low temperatures. Moreover, the hardness of nickel films' ferromagnetic loop of anomalous Hall effect increases and then decreases, which reflects the nickel film's transtion from IMA at 24 nm to PMA at 8 nm, before reverting to IMA at 3 and 2 nm at low temperatures, exhibiting strong thickness dependence of the magnetic anisotropy. Additionally, a butterfly magnetoresistance arises due to the anisotropic magnetoresistance effect, which becomes the sharpest for the 8 nm-thick nickel film with PMA and also supports the thickness-dependent magnetic transitions. Furthermore, the square ferromagnetic loop changes from a hard to a soft one at rising temperature, revealing a PMA-to-IMA shift as the system is heated above 125 K, co-evidenced by both the magnetic hysteresis loops and the butterfly magnetoresistance. These fickle magnetic transformations have been systematically unveiled by using transport as a probe and provide valuable guidance to PMA engineering for designing nickel-based spintronics.

## 2. Experimental details

The nickel films were deposited onto the 10×10 cm $SiO_2$/Si substrates using a high-vacuum electron beam evaporation system. X-ray diffraction (XRD) measurements were performed to confirm the nickel films' crystallinity. As shown in Fig. S1, the nickel films are in a polycrystalline status and host a cubic structure with space group Fm3m (225) and a lattice constant of 0.35238 nm (matched with the standard PDF#04-0850). The evaporation rate was approximately 0.01 nm/s. By controlling the evaporation time, we got the samples with different thicknesses of 24, 8, 3, and 2 nm respectively. Afterward, gold wires were used to connect the sample to the electronic chip via silver paste, completing the sample fabrication for electrical transport measurements. We carried out transport measurements using a resistivity probe operated inside a cryostat, reaching temperatures as low as 2 K and magnetic fields up to 9 T. The Keithley 6220 was used to apply the DC current, and two Keithley 2182 were adopted to measure longitudinal and transverse resistance simultaneously.

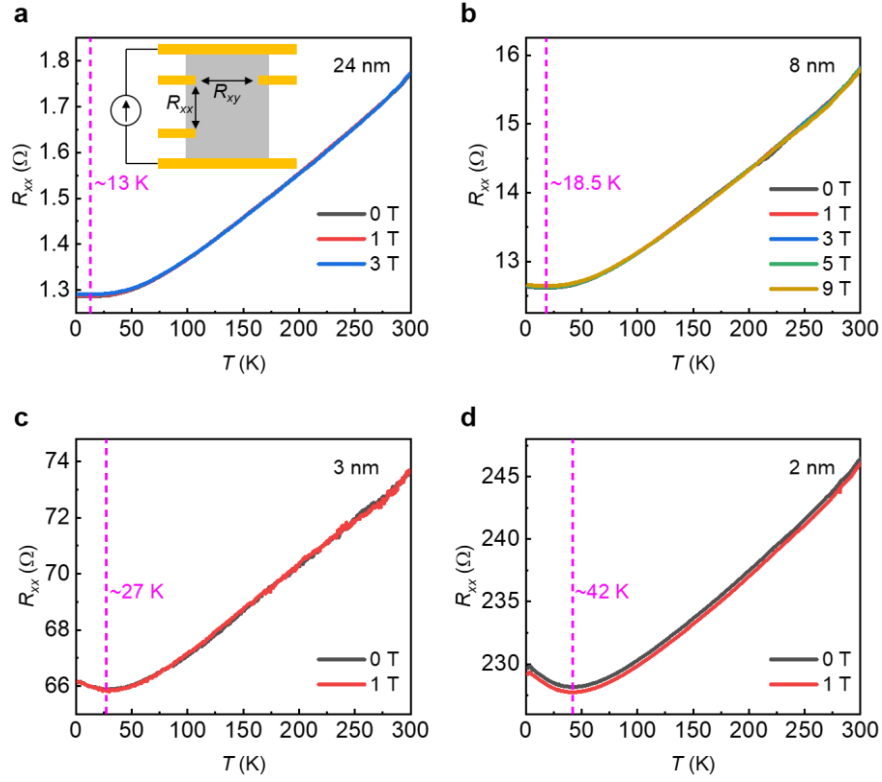

**Figure 1.** Temperature-dependent longitudinal resistance. (a-d) Longitudinal resistance $R_{xx}$ as a function of temperature $T$ for nickel films with thickness of 24 nm (a), 8 nm (b), 3 nm (c), and 2 nm (d), respectively. The $R_{xx}$-$T$ curves are independent of magnetic fields. A residual resistance emerges at ~13 K for 24 nm-thick sample, and a resistance upturn develops at ~18.5, ~27, ~42 K for 8, 3, 2 nm-thick samples, respectively (indicated by pink dashed lines). The inset in (a) illustrates the transport measurement geometry, where current is applied from the top to bottom electrodes on nickel film. The longitudinal (transverse) resistance $R_{xx}$ ($R_{xy}$) is monitored with regard to the two electrodes indicated by the double-headed arrow correspondingly.

## 3. Results and discussions

### 3.1 Temperature-dependent longitudinal resistance

We fabricated four nickel films with thicknesses of 24 nm, 8 nm, 3 nm, and 2 nm, which fall within the typical thickness range of magnet layers used in practical magnetic devices. Figure 1 displays the temperature dependence of the longitudinal resistance ($R_{xx}$-$T$ curve) for each of the samples. The $R_{xx}$-$T$ curves for all samples show a prototypical metallic behavior, with decreasing $R_{xx}$ as $T$ decreases, and are insensitive to external magnetic fields. Additionally, a residual resistance appears below ~13 K for the 24 nm-thick nickel film (Fig. 1a) and gradually develops a significant resistance upturn at around 18.5 K, 27 K, and 42 K for the 8 nm, 3 nm, and 2 nm-thick samples, respectively. Since this resistance upturn is almost independent of magnetic fields, it cannot be attributed to the Kondo effect[33, 34] that significantly suppresses the resistance upturn at high magnetic fields. Instead, it is more likely due to the strengthening of electron-electron interaction[35] in thinner films.

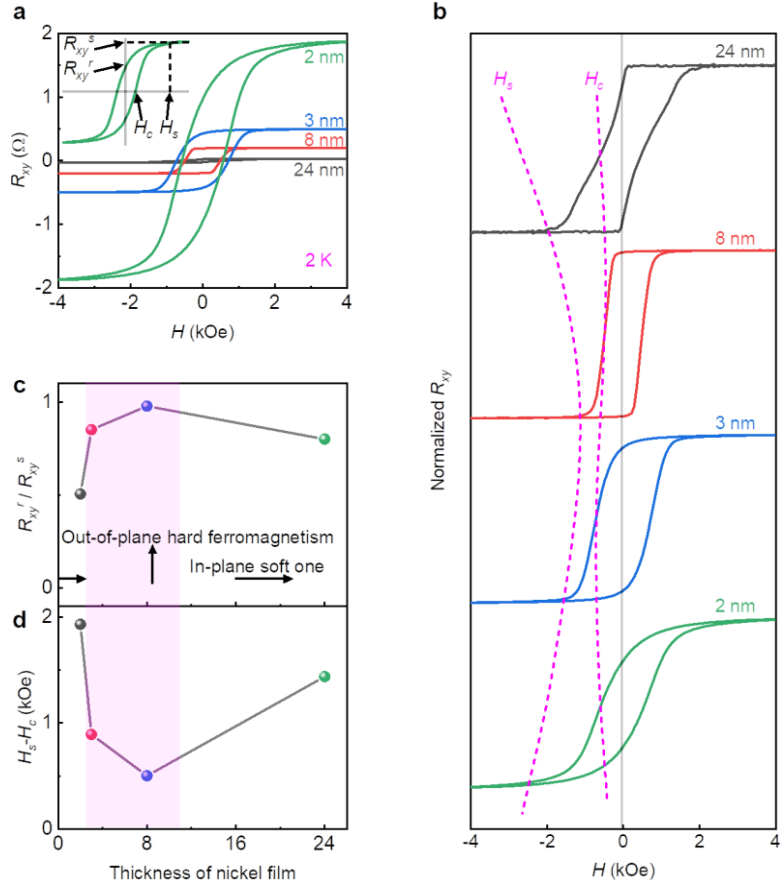

**Figure 2.** Thickness-dependent magnetism. (a) Transverse resistance $R_{xy}$ as a function of magnetic field for all the samples at the lowest temperature of 2 K. It shows a typical ferromagnetic hysteresis loop due to the anomalous Hall effect. The inset illustrates the four parameters that can be derived from the $R_{xy}$-$H$ curve. (b) Normalized $R_{xy}$-$H$ curve for each sample. The hysteresis loop of the 8 nm-thick sample is much more rectangular than other slanted loops. Although $H_c$ is not much varied across individual samples, $H_s$ changes significantly with its minimum at 8 nm (indicated by the two pink dashed curves). (c-d) $R_{xy}^r/R_{xy}^s$ (c) and $H_s$-$H_c$ (d) as a function of thickness. $R_{xy}^r/R_{xy}^s$ reaches a maximum of 1 and $H_s$-$H_c$ reaches a minimum at 8 nm, demonstrating the hard ferromagnetism with perpendicular magnetic anisotropy. It indicates the IMA-PMA-IMA transition as thickness reduces from 24 to 2 nm, which is also supported by previous investigations by other methods[36, 37].

*3.2 Thickness-dependent magnetism*

Figure 2a depicts the transverse resistance as a function of the out-of-plane magnetic field ($R_{xy}$-$H$ curve) for nickel films at the lowest temperature of 2 K. The curve displays a distinct hysteresis loop, resulting from the anomalous Hall effect in a ferromagnet, where the transverse resistance $R_{xy}$ is linearly proportional to the magnetization $M_z$. From a ferromagnetic hysteresis loop, four physical parameters can be derived: $R_{xy}^s$, $H_s$, $R_{xy}^r$, and $H_c$ (as shown in the inset of Fig. 2a). These parameters correspond to the saturated magnetization at high magnetic fields, the magnetic field required for magnetization saturation, the remnant magnetization when the magnetic field returns to zero, and the coercive field at zero magnetization, respectively.

In Figure 2b, the transverse resistance $R_{xy}$ is normalized by $R_{xy}^s$, allowing for a clearer comparison of the shapes of the hysteresis loops and the evolution of the physical parameters $H_s$ and $H_c$. As the film thickness decreases from 24 nm to 2 nm, the hysteresis loop becomes more rectangular at 8 nm and then returns to a heavily skewed one. Importantly, the magnetic field gap between $H_s$ and $H_c$ narrows at 8 nm before widening again, indicating a transition from IMA to PMA and back to IMA. The hysteresis loop's hardness is characterized by the $R_{xy}^r/R_{xy}^s$ ratio and the $H_s$-$H_c$ (shown in Fig. 2c and d, respectively). The 8 nm-thick nickel film exhibits a perfect $R_{xy}^r/R_{xy}^s$ ratio of 1 and the minimum $H_s$-$H_c$, both of which demonstrate its PMA nature in contrast to the IMA trait of the other three samples.

Strictly speaking, the electrical transport method itself is not sufficient to completely determine PMA and IMA. Specifically the $R_{xy}$-$H$ curve can only show the hardness of a ferromagnet, reflecting the relative PMA change but showing less information on IMA. Fortunately, previous investaigations[36, 37] have adopted different methods to tackle the same problem and got the same conclusion: on increasing thickness, the IMA dominants below 7 monolayers (~2.5 nm), then switches to the PMA, and finally returns to the IMA above 28 monolayers (~10 nm). The PMA peaks at ~7 nm.

Based on our understanding and previous exclusive literature[36-38], this magnetic anisotropy transition can be explained using the effective anisotropy energy scenario. The system contains three primary forms of magnetic anisotropy energy: Neel-type surface anisotropy $K_{\text{surface}}$, magnetocrystalline anisotropy $K_u$ by strain and tetragonal distortion as the perpendicular uniaxial anisotropy, and shape anisotropy $K_{\text{shape}}$ associated with demagnetization energy. In the case of the 2 and 3-nm-thick samples, the surface anisotropy $K_{\text{surface}}$ is exceptionally high and determines the spin easy axis to be in-plane. As the thickness increases to 8 nm, $K_{\text{surface}}$ is significantly reduced, and $K_u$ becomes dominant, leading to an out-of-plane easy axis. For very thick nickel films, such as the 24 nm-thick sample, the shape anisotropy $K_{\text{shape}}$ outweighs the other two due to the strong demagnetization effect along the out-of-plane direction, resulting in an in-plane easy axis as frequently seen for most ferromagnetic films. Since it is beyond the experimental capability of our transport tools, we cannot judge precisely the exact source of the $K_u$. Nevertheless, the polycrystalline structures in Fig. S1 provide possibilities to bear the $K_u$. Note that this effective anisotropy scenario has been well established by individual works[36-38], and our transport studies are consistent with this magnetic transition physical picture by using transport as a probe. This thickness-dependent PMA transition also implies that one should pick up the optimal thickness of a ferromagnet for stabilizing the PMA when combing the ferromagnet with heavy-atom compounds[39, 40] for spin-orbit torque spintronics. Concretely for nickel film, the optimal thickness is ~8 nm to stabilize the PMA for spin-orbit torque application, hosting the squarest ferromagnetic loop of highest hardness with $R_{xy}^r/R_{xy}^s$~1 and minimum $H_s$-$H_c$. This piece of information is valuable experimental guidance for current-driven spintronics based on nickel films that our work can contribute.

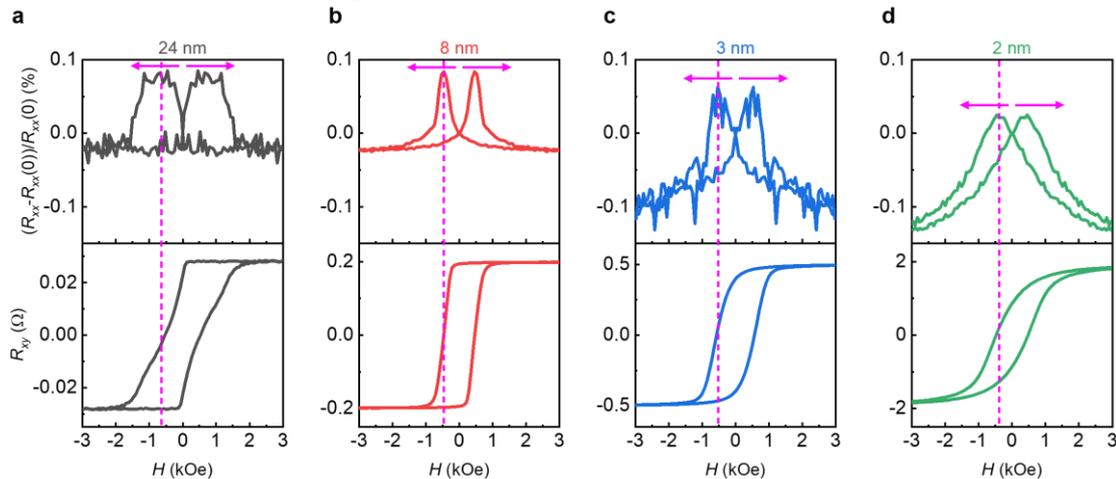

**Figure 3.** Butterfly magnetoresistance. (a-d) Longitudinal magnetoresistance ratio $(R_{xx}-R_{xx}(0))/R_{xx}(0)$ and $R_{xy}$ as a function of the magnetic field at 10 K, for 24 (a), 8 (b), 3 (c), and 2 (d) nm-thick samples, respectively. Here, $R_{xx}$ indicates the longitudinal resistance at magnetic field $H$ while $R_{xx}(0)$ represents the longitudinal resistance at zero magnetic field. Butterfly magnetoresistance emerges near ferromagnetic transition, and peaks at the coercive field (indicated by pink dashed line) while sweeping the magnetic field (sweeping direction indicated by pink arrows). Butterfly magnetoresistance peak sharpens most for the 8 nm-thick PMA ferromagnet but much broadens for all other IMA samples with thicknesses of 24, 3, and 2 nm.

*3.3 Butterfly magnetoresistance*

In addition to the ferromagnetic anomalous Hall effect discussed earlier, the longitudinal magnetoresistance is also probed as a function of the magnetic field. The upper panels in Figure 3a-d present the magnetoresistance ratio $(R_{xx}-R_{xx}(0))/R_{xx}(0)$ for each sample at 10 K, where $R_{xx}$ indicates the longitudinal resistance at magnetic field $H$ while $R_{xx}(0)$ represents the longitudinal resistance at zero magnetic fields. The pink left-headed (right-headed) arrow indicates the magnetic field sweeping from positive to negative (from negative to positive) direction. It can be observed that butterfly magnetoresistance emerges during the ferromagnetic transition by magnetic field and peaks around the coercive field $H_c$ (indicated by the pink dashed line) for all our samples.

A similar butterfly magnetoresistance was also observed in the van der Waals ferromagnet Fe$_3$GeTe$_2$[41] due to the in-plane anisotropic magnetoresistance effect and the modulated magnon scattering. We propose a similar mechanism for the nickel film, taking the 8 nm-thick sample as an example. While sweeping the magnetic field from positive to zero field, the magnetoresistance just slightly increases due to the enhanced magnon population and corresponding magnon scattering. When the magnetic field keeps swept toward the negative direction, more spins will be altered from out-of-plane to in-plane direction. Such spin reorientation boosts the probability of spins aligned parallel to the current direction, causing continually enhanced magnetoresistance via the famous in-plane anisotropic magnetoresistance effect. The magnetoresistance peaks at the coercive field $H_c$, where all the spins are located in-plane in principle. If the magnetic field is further swept to negative high values, spins will be reoriented from in-plane to downward out-of-plane direction, which is vertical to the current and thus leads to lower magnetoresistance. Finally, the in-plane anisotropic magnetoresistance effect produces the butterfly-shaped magnetoresistance. Such a scenario is also evidenced by its spin evolution captured by the ferromagnetic loop of anomalous Hall effect (the lower panels in Fig. 3a-d).

Notably, the butterfly peak sharpens most for the 8 nm-thick sample with PMA, and broadens for the three other samples with IMA, following the $H_s$-$H_c$ evolution on thickness in Fig. 2d. Meanwhile, the difference between $R_{xx}$(peak) and $R_{xx}$(0), i.e., ($R_{xx}$(peak)- $R_{xx}$(0))/$R_{xx}$(0) display a rough trend of 8 nm > 24 nm ≈ 3 nm > 2 nm, consistent with the $R_{xy}^r/R_{xy}^s$-thickness dependences in Fig. 2c. These two consistencies not merely prove the thickness-dependent IMA-PMA-IMA magnetic transition once again, but also support the in-plane magnetoresistance effect as the dominant explanation for the butterfly-shaped magnetoresistance. Butterfly magnetoresistance represents a common phenomenon in 3d transition metals, which has been ascribed to the ordinary magnetoresistance effect (high-filed linear part) and the anisotropic magnetoresistance effect (butterfly peak part)[42]. Here we are just using anisotropic magnetoresistance effect-induced butterfly magnetoresistance ($R_{xx}$-$H$) as another complementary support for the ferromagnet's harness change in thickness, crosschecked with the ferromagnetic loop by anomalous Hall effect ($R_{xy}$-$H$) simultaneously.

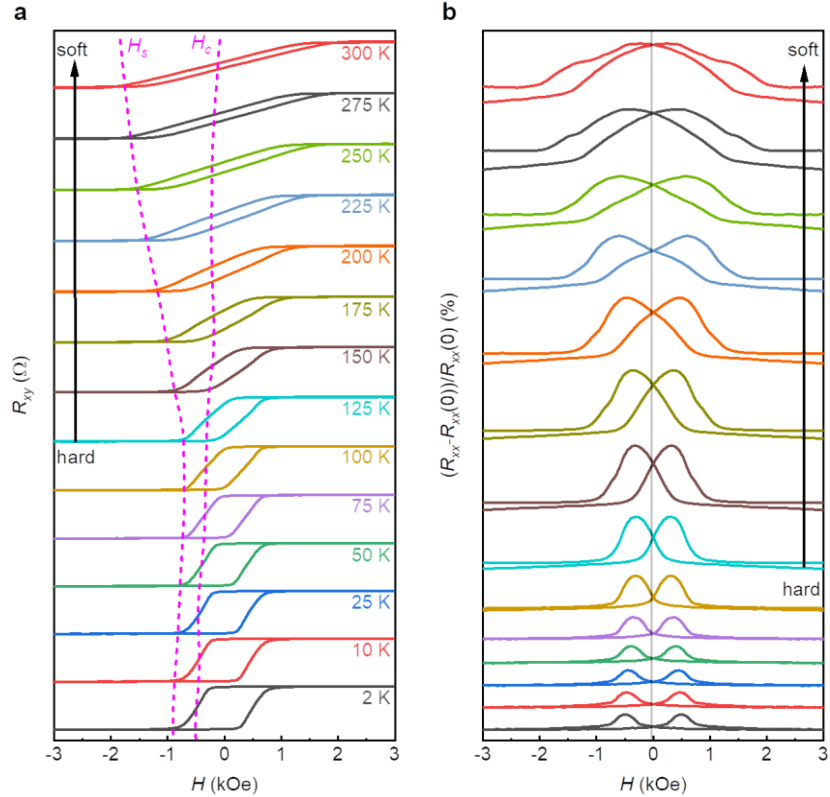

**Figure 4.** Temperature-dependent magnetic transition. (a) $R_{xy}$-$H$ curves at various temperatures for the 8 nm-thick sample. The dashed lines highlight the $H_s$ and $H_c$ evolution. From 2 to 100 K, $H_s$ and $H_c$ don't change much and the $R_{xy}$-$H$ curves exhibit well-behaved hard ferromagnetism. From 125 to 300 K, $H_s$ significantly increases with increasing temperature, featuring a hard-to-soft transition. (b) Longitudinal magnetoresistance ratio ($R_{xx}$-$R_{xx}$(0))/$R_{xx}$(0)-$H$ curves at various temperatures. The butterfly magnetoresistance peak is little changed up to 100 K, but greatly broadens from 125 to 300 K. These behaviors indicate the PMA-to-IMA transition while increasing temperature above 125 K.

### 3.4 Temperature-dependent magnetic transition

We further inspect the temperature dependence of magnetism, primarily on the 8 nm-thick nickel film. Figure 4a illustrates the ferromagnetic $R_{xy}$-$H$ loop at various temperatures ranging from 2 to 300 K, where the saturation field $H_s$ and coercive field $H_c$ are highlighted by pink dashed curves. Between 2 and 100 K, the 8 nm-thick nickel film preserves a well-behaved ferromagnet of PMA with nearly unchanged $H_s$ and $H_c$. However, unexpectedly, as the temperature is increased above 125 K, although the coercive field $H_c$ shows little variation, the saturation field $H_s$ increases roughly linearly. Such increased $H_s$ results in a heavily slanted hysteresis loop instead of the previous rectangular one.

This temperature-dependent transition behavior is in sharp contrast to standard hard ferromagnets like $Fe_3GeTe_2$[19, 20]. For $Fe_3GeTe_2$, the rectangular shape of the ferromagnetic loop preserves from the base temperature to the Curie temperature, and $H_s$ along with $H_c$ decreases simultaneously and monotonically until vanishing. In contrast, the nickel sample exhibits a transition behavior that is similar to the PMA-to-IMA transition shown in Fig. 2. It is more akin to the spin reorientation transition in strange ferromagnets such as $Fe_4GeTe_2$[43], where occurs the temperature-induced transformation from PMA to IMA. In $Fe_4GeTe_2$[43], this spin reorientation transition was explained as the consequence of the small effective magnetic anisotropy ($K_{eff}$), which originates from two competing contributions: the magnetocrystalline anisotropy ($K_u$), favoring the out-of-plane anisotropy, and the shape anisotropy ($K_{shape}$), favoring the in-plane anisotropy. They usually follow different temperature dependences, roughly $K_u \sim M_s^3$ and $K_{shape} \sim M_s^2$, which leads to the temperature-driven spin reorientation transition. The situations in our nickel film system are similar: for the 8 nm-thick nickel film, basically $K_{surface}$ is rather small; $K_u$ competes with $K_{shape}$ and overwhelms it at low temperature to stabilize PMA at low temperature. However, at high temperatures above 125 K, $K_u$ reduces much faster than $K_{shape}$, eventually leading to the PMA to IMA transition at high temperatures.

As another piece of independent evidence shown in Fig. 4b, the butterfly magnetoresistance peak gets significantly broadened above 125 K, and the ($R_{xx}$(peak)- $R_{xx}$(0))/$R_{xx}$(0) gradually reduces to nearly zero from 125 to 300 K. Therefore, based on these experimental findings and detailed analyses, we can conclude that the transition from hard to soft magnetism above 125 K is mainly ascribed to the temperature-induced PMA-to-IMA transition. This spin reorientation transition warns us that one should take extra care like annealing, etc., to enhance its PMA if a nickel film is adopted for high-temperature applications.

## 4. Conclusions

In summary, we have reported comprehensive electrical transport studies on the ferromagnetic nickel films below tens of nanometers thick. As the thickness decreases, the longitudinal residual resistance evolves into a growing resistance upturn from a residual resistance plateau at low temperatures. In addition, we have discovered the IMA-PMA-IMA transition, and observed a butterfly magnetoresistance near the ferromagnetic transition that peaks at the coercive field and sharpens mostly for the PMA nickel sample. Furthermore, the PMA nickel ferromagnet undergoes a PMA-to-IMA transition above 125 K. All of these results demonstrate the highly versatile properties of nickel film in every aspect and provide guidance for engineering nickel-based ferromagnets in magnetic device applications, substantially pushing forward the nickel-based spintronics.

## Data availability statement

All data that support the findings of this study are included within the article.

## Acknowledgements

We acknowledge Prof. Je-Geun Park for his helpful discussions and support. The work at CQM and SNU was supported by the Leading Researcher Program of the National Research Foundation of Korea (Grant No. 2020R1A3B2079375), Samsung Science & Technology Foundation (Grant No. SSTF-BA2101-05), and the Samsung Advanced Institute of Technology. This work was also supported by the National Natural Science Foundation of China (No. 62104703), and The Pearl River Talent Recruitment Program (No. 2019ZT08X639).

**Supporting Information:**

**Supporting Figures**

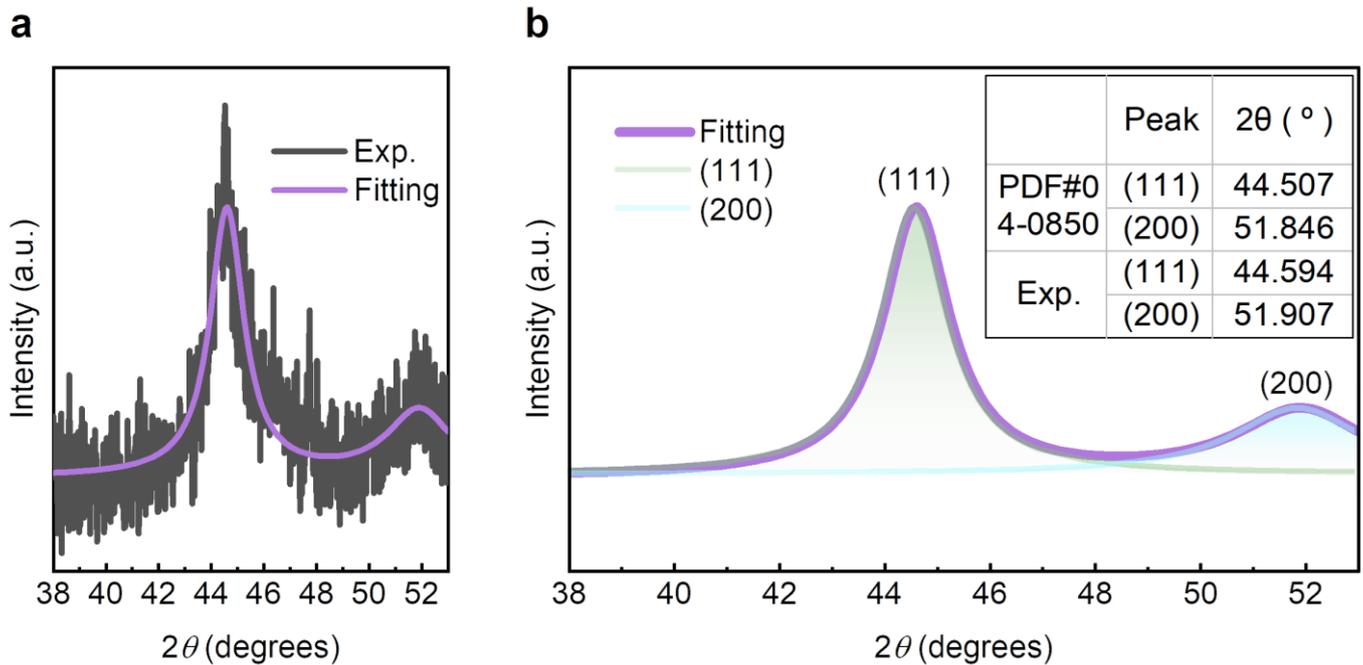

**Figure S1.** Powder XRD measurement and analysis of the nickel film. (a) Powder XRD result of the nickel film on SiO$_2$/Si substrates. The black and purple lines indicate the experimental spectra and the peak fitting result. The standard PDF#04-0850 of cubic nickel (Space group: Fm3m(225)) reveals the two main peaks to be (111) and (200). Therefore we only focus on this spectra region to avoid strong disturbance from the silicon substrates. (b) Fitting curve and separated experimental peaks (111) and (200), highlighted in purple, green, and blue, respectively. The inset table represents the comparison between the standard PDF#04-0850 and our experiment, which match with each other.